\documentstyle[twocolumn,prl,aps,epsfig]{revtex}
\newcommand{\be}{\begin{eqnarray}}
\newcommand{\ee}{\end{eqnarray}}

\begin{document}
\draft

\title{Single and Paired Point Defects in a 2D Wigner Crystal}

\author{Ladir C\^andido, Philip Phillips, and D. M. Ceperley}
\vspace{.05in}

%
\address
{Loomis Laboratory of Physics\\ and NCSA,
University of Illinois
at Urbana-Champaign\\ 1100 W. Green St., Urbana, IL, 61801-3080}

%

\address{\mbox{ }}
\address{\parbox{14.5cm}{\rm \mbox{ }\mbox{ }
Using the path-integral Monte Carlo method, we calculate the
energy to form single and pair vacancies and interstitials in a
two-dimensional Wigner crystal of electrons. We confirm that the
lowest-lying energy defects of a 2D electron Wigner crystal are
interstitials, with a creation energy roughly 2/3 that of a
vacancy. The formation energy of the defects goes to zero near
melting, suggesting that point defects might mediate the melting process. 
In addition, we find that the interaction between defects is strongly
attractive, so that most defects will exist as bound pairs.}}
\address{\mbox{ }}
\address{\mbox{ }}%


\maketitle

At sufficiently low densities, the ground state of a 2D electron
gas is expected to be a Wigner crystal\cite{wigner}. Quantum Monte
Carlo calculations\cite{ceperley} predict Wigner crystal formation
for $r_s>37$ at zero temperature where $r_s= a_{W}/a_0 $ is the
dimensionless parameter characterizing the ratio of potential to
kinetic energy. QMC calculations also predict that disorder
stabilizes the solid phase relative to that of the liquid
\cite{chui} thereby shifting the melting boundary to higher
densities, $r_s\approx 10$.  In light of recent experimental
observations\cite{k1,k2,k3,pop1,pepper,shahar,yoon,mills,papa} of
a new conducting phase in 2D in the density range $10<r_s<37$,
there is renewed interest in the properties of a dilute electron
gas close to the Wigner crystal melting boundary.  Yoon, et.
al.\cite{yoon} have observed a transition at $r_s\approx 40$ which
they attribute to the melting of a Wigner crystal.

In this work, we calculate the energies of point-like defects in a
clean 2D electron Wigner crystal.  The motivation for the study of
defects is two-fold.  First, localized defects are present in a
finite concentration at any nonzero temperature (they have even
been speculated to exist at zero temperature a super solid).
Secondly, the melting process in 2D can be influenced or even
determined by defect formation\cite{melting}. We investigate the
energy of two kinds of point defects: a vacancy at one of the
lattice sites or an interstitial centered in one of the triangular
unit cells.  We define $N_{\rm def}$ ( the defect index) as the
number of electrons minus the number of lattice sites. Using
Path-Integral Monte Carlo (PIMC), we compute the energy to
introduce defects into an N-electron crystal with $-2\le N_{\rm
def}\le 2$. We find that the lowest lying defect excitations are
centered interstitials. Similar results have been obtained by Jain
and Nelson~\cite{jn} in hexagonal columnar crystals and by
Cockayne and Elser\cite{ce} in the 2D Wigner crystal. We then show
that for $N_{\rm def}=\pm 2$, the defect creation energy increases
as the defects are pulled apart.  This indicates that the ground
state for two defects is a bound pair.

The system being simulated is composed of electrons confined to
two dimensions (2D) and interacting through a repulsive $1/r$
potential and immersed in a positive background with the same
density. The Hamiltonian is
\be
H = {1 \over r_{s}^{2}}\sum_{i=1}^{N}\nabla^2_{i} + {2 \over
r_{s}}\sum_{i<j}^{N}{1 \over \vert {\bf r}_{i} - {\bf r}_{j} \vert } 
+V_0,
\ee
where energy is in units of the Rydberg $\hbar^2/2ma_{0}^{2}$,
$a_0$ is the effective Bohr radius and lengths are in units of the
Wigner-Seitz radius $a_W=(\pi \rho)^{-1/2}$ and $V_0$ a constant.
To understand the role of defects in well-characterized electron
crystals, we focus on systems with $r_s \geq 40$.

At large values of $r_s$, the exchange contributions to the energy
are small\cite{ceperley,zhu95}. For sufficiently large $r_s$,
$r_s>75$, we consider distinguishable electrons and neglect
antisymmetry but include it for $r_s \leq 75$. However, to keep
the system stable, we forbid particle exchange by enforcing the
condition $|{\bf r}_i-{\bf s}_i|< 1.1 a$ where $a$ is the nearest
neighbor distance and ${\bf s_i}$ is the $i^{th}$ lattice site.
Such a ``tether'' is realistic because in a quantum crystal
exchanges are rare and the wave function is peaked around the
lattice sites. We have verified the independence of our results by
varying and removing this constraint. In these calculations we
assumed a hexagonal lattice, shown to be the stable
structure\cite{zhu95}.

To study defect formation in a Wigner crystal, we use the PIMC
method\cite{pol}. In this method, the density matrix
\be\label{density} \rho(R_{0},R_{M};\beta)&=&\int...\int
dR_{1}dR_{2}... dR_{M-1}\rho(R_{0},R_{1};\tau)\nonumber\\
&&\rho(R_{1},R_{2};\tau)...\rho(R_{M-1},R_{M};\tau). \ee for the
quantum system is evaluated by sampling paths:
$\{R_{0},R_{1}...R_{M-1},R_{M}\}$, and $R_{k}= \{{\bf
r}_{1,k},...{\bf r}_{n,k}\}$ and ${\bf r}_{i,k}$, a bead, is the
position of $i$th electron in the $k$th time slice and
$\tau=\beta/M$ with $\beta=1/k_{B}T$. The action
$-\ln(\rho(R_{M-1},R_{M};\tau))$ is evaluated by first splitting
the potential into a long-range part and a short-ranged part. The
long-ranged part (which is slowly varying) is handled in the
primitive approximation, while the short ranged action is the
exact pair action of two electrons. To evaluate the
2NM-dimensional integrals in Eq.(\ref{density}), we employ the
Metropolis bisection sampling technique\cite{rmp}.
Ewald\cite{ewal} sums are used to calculate the long ranged
potential energy and pair action\cite{nat}.

At densities with $r_s\leq 75$ we used restricted path
integrals\cite{cep92} to account for Fermi statistics and
stabilize the crystal against melting. Only paths which remained
in the positive region of the Slater determinant were allowed.
Such a restriction is exact if the nodal surfaces of the trial
density matrix are correct. For the nodes we used a Slater
determinant of Gaussian orbitals $\exp(-c({\bf r}_i-{\bf s}_j)^2)$
with ${\bf s}_i$ the lattice sites and $c$ taken from \cite
{cep78} with a ferromagnetic spin arrangement. Calculation of the
multiple exchange Hamiltonian necessary to determine the ground
state magnetic ordering is in progress but WKB calculations
indicate that ferromagnetism is most likely\cite{roger}.

\begin{figure}
\begin{center}
\epsfig{file=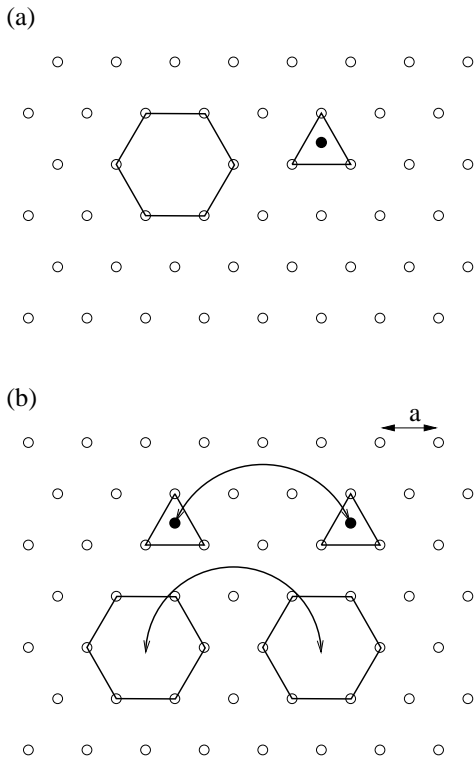, height=10.0cm}
\vspace{.2in} \caption{a) Hexagonal 2D Wigner crystal with a
single 6-coordinated vacancy and a 3-fold coordinated centered
interstitial (CI) defect.  b) Pair vacancy and CI defects. }
\label{fig1}
\end{center}
\end{figure}

The simulations were performed at fixed temperature, area and number
of particles. The energy to create $N_{\rm def}$ defects (see Eq.
(\ref{edef})) in a system of $N$ particles and $N_{l}=N$ lattice
sites with a density of $\rho=N/A$ is
\be
\label{edef} \Delta E_{\rm def} =\left[ e(N+N_{\rm def})-
e(N)\right ](N+N_{\rm def}), \ee where $e(n)$ is the energy per
electron for a system containing $n$ electrons, a density $\rho$
and $N_{\ell}$ lattice sites. The densities and temperatures
studied were $r_s =$40, 50, 75, 100, and 200 and T=5.0, 2.5 and
1.25$\times 10^{-5}$Ry. We used $120$ and $340$ lattice sites. We
performed separated PIMC calculations with different number of
particles and lattice sites rather than more efficient procedures
where particles are inserted or removed\cite{gillan}. Such
differential procedures are difficult because of the combination
of large relaxation of the lattice, the very large zero point
motion and the antisymmetry. Our calculations are very
time-consuming as the system gets larger because one needs high
accuracy to obtain the energy difference but the direct method
allows better control over the systematic error.

To illustrate the types of defects of interest, we depict in Fig.
\ref{fig1} vacancy and centered-interstitial (CI) defects in a 2D
hexagonal lattice. The initial configuration of the vacancy and CI
defects have 6-fold and 3-fold coordination, respectively. In PIMC
all particles are dynamical variables, so that the surrounding
lattice relaxes and may change the overall symmetry of the defect.
However, the constraint makes such relaxation local. We find that
the defect energy depends weakly on temperature. The following
discussion concerns the defect energies for the temperature
1.25$\times10^{-5}$Ry. This is significantly less than the melting
temperature and the Debye temperature.

\begin{figure}
\begin{center}
\epsfig{file=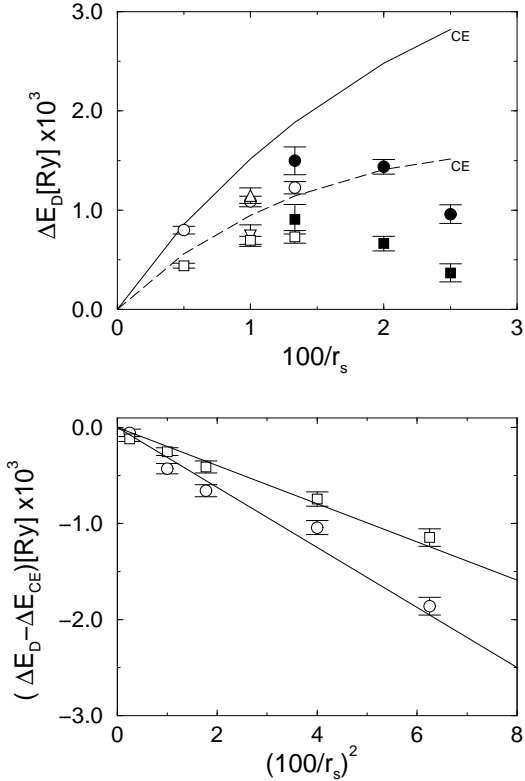, height=10.5cm} \caption{ Formation energy
(top figure) for vacancy (circle) and CI (square) defects as a
function of $1/r_s$ at a temperature of 1.25$\times 10^{-5}$Ry for
a system of 120 lattice sites (open and solid simbols are for
Boltzman and Fermi statistics respectively). The triangles are
results with 340 lattice sites. The lines (solid for vacancy and
dashed for interstitial) are from harmonic
calculations\protect\cite{ce}. The anharmonic energy is shown on
the bottom panel.} \label{fig2}
\end{center}
\end{figure}

Fig. \ref{fig2} demonstrates that the formation energy for an
interstitial is consistently lower than creation energy for
vacancies in 2D Wigner crystals at all densities. At $r_s=75$ we
find that $\Delta E_{\rm CI}\approx 0.65 \Delta E_{\rm vac}$.
Simulations with 340 sites give agreement with the results for the
smaller lattice. Jain and Nelson\cite{jn} and Cockayne and
Elser\cite{ce} have obtained a similar result.

\begin{figure}
\begin{center}
\epsfig{file=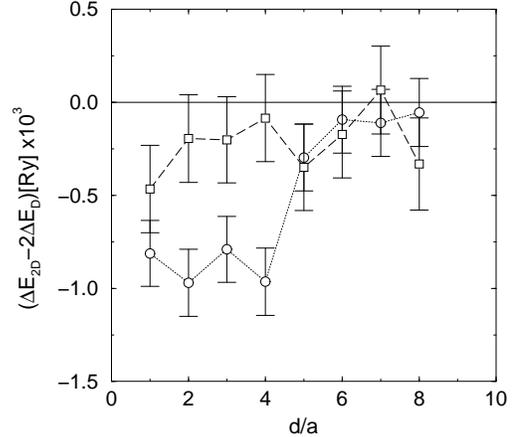, height=6.0cm} \caption{Binding energy of
pair defects at $r_s$=100 and T=1.25$\times 10^{-5}$Ry, computed
for systems of 340 lattice sites. The geometry used to study such
defects is shown in Fig.(1b).  Circles are for vacancies and
squares for interstitials. } \label{fig3}
\end{center}
\end{figure}

\begin{figure}
\begin{center}
\epsfig{file=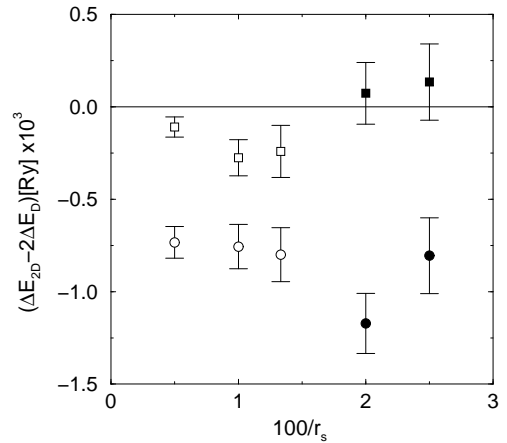, height=6.0cm} \caption{Binding energy of
pair defects  separated by a single lattice site as a function of
$1/r_s$ at T=1.25$\times 10^{-5}$Ry, computed for systems of 120
lattice sites. Interstitials correspond to squares and vacancies
to circles. } \label{fig4}
\end{center}
\end{figure}

For a Coulomb system without fermion exchange, the energy of a
defect can be expanded as \be\label{expansion} E_D = c_1r_s^{-1}
+c_{3/2} r_s^{-3/2} + c_2 r_s^{-2} \ldots \ee The first term is
the static potential energy of the defect, the second the harmonic
energy of the defect. Cockayne and Elser\cite{ce} have done exact
calculation of $c_1$ and $c_{3/2}$. Shown in the lower panel of
Fig. \ref{fig2} is the anharmonic contribution defined as the
excess energy beyond the harmonic calculation. Anharmonic effects
lower the defect energies to approximately half the harmonic value
at $r_s=50$. Cockayne and Elser find that the defect energy
vanishes for CI defects at $r_s=15\pm 1$ and $r_s=9\pm 1$ for
vacancies. We obtain $c_2(vacancy) = -2.8 \pm 0.2$ and $c_2(CI) =
-1.85\pm 0.15$.  From these values we estimate that $E_D$ vanishes
for interstitials at $r_s=35 \pm 2$ and $r_s=29 \pm 2$ for
vacancies.  The proximity of the vanishing of the interstitial
creation energy to the melting density is highly suggestive that
interstitial defects play a role in the melting process. It is
possible that the assumption of the ferromagnetic spin arrangement
is stabilizing the crystal with respect to interstitials for  $35
\leq r_s \leq 40$. We have not considered the role of other
defects such as dislocations.
We have also investigated the properties of pair defects as a
function of the spacing between the defects. The geometry used for
the study of the interaction energy between two defects is shown
in Fig. \ref{fig1}b. Shown in Fig. \ref{fig3} are the pair
binding energies for vacancies and CI defects:
\be
E_{bp}=\Delta E_{\rm def}(N_{\rm def}=2)-2\Delta E_{\rm
def}(N_{\rm def}=1) \ee  as a function of the separation between
the defects. As expected, for large separations the binding energy
goes to zero. For all separations studied, the pair-binding energy
is negative, indicating an attraction between defects.  As is evident from 
Fig. \ref{fig4}, the interstitial
pair-binding becomes slightly positive at $r_s\approx 50$
indicating a possible weak-unbinding near the melting transition.
No such behaviour was observed for vacancies, however. In two
dimensions, any attraction is sufficient for the existence of a
bound state of two defects. At $r_s=100$, the binding energy for
interstitials pairs is roughly $50K$, for vacancy pairs 125K and
the classical melting temperature is only 24K (assuming the
effective mass of the electrons and the dielectric constant are
unity). The symmetry of this paired state will be determined by
the magnetic order of the lattice and the relative exchange
frequencies of the defects. For a 3D Wigner crystal,
Moulopoulos and Ashcroft\cite{ma} have proposed that a paired Wigner state is a
possible intermediary to melting.  While our calculations illustrate that
interstitials are strongly bound, they do not indicate that a lattice
of such paired defects is stable.

\begin{figure}
\begin{center}
\epsfig{file=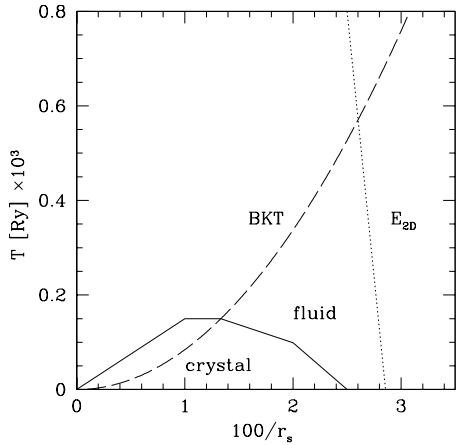, height=7.5cm} \caption{Phase diagram. The
solid line is an estimate of the stable region of the Wigner
crystal. The dotted line indicates the energy of the paired
interstitial, the dashed line the condition for the BKT transition,
$E_{2D}<0.84/r_s^2$. } \label{fig5}
\end{center}
\end{figure}

As seen in Fig. \ref{fig5} the binding energy of the defect is
much greater than the melting temperature, hence point defects are
mainly interstitial pairs, which are expected to obey Bose
statistics.
Let us consider whether one could have a Bose
condensation of defect pairs. The defect pairs are thermally
activated so that their density is $\rho_{2D}=\rho_0
\exp(-E_{2D}/k_B T)$ with $\rho_0 =1/\pi$ and $E_{2D}$ the
formation energy of a pair defect. The
Brezinski-Kosterlitz-Thouless (BKT) transition occurs at a density
$\rho_{2D}\lambda^*/k_B T \approx 0.28$ (this is scaled from the
helium data\cite{rmp}). A solution to these two equations exists
provided that $E_{2D} < 0.42 \lambda^*$. Here $\lambda^*
=\hbar^2/(2m^*) < 2/r_s^2$  since the pair defects are heavier
than two free electrons. Hence a necessary condition for the
existence of a super-solid phase is that $E_{2D} < 0.84/ r_s^2$.
This condition is shown in Fig. \ref{fig5}. Our results indicate
that the supersolid transition could only occur in the Wigner
crystal very near melting and is likely to be precluded by the
unbinding of pairs.

As mentioned above, we find that in a 2D Wigner crystal most
defects will be bound pairs of interstitials. The conductivity is
likely to be dominated by defect transport. This could explain the
extreme sensitivity of the conductivity in the insulating
phase\cite{k3} to an in-plane magnetic field which would cause
unbinding of singlet pairs. Further simulations are needed to
decide whether pairing persists in the melted phase as has been
proposed by one of us\cite{nature}.

\section*{Acknowledgments}
PP acknowledges stimulating discussions with Antonio Castro-Neto
and partial funding from DMR98-12422. LC acknowledges useful discussions and computational help
with Dr. G. Bauer and Dr. B. Militzer and support by Fundac\~ao de
Amparo \`a Pesquisa do Estado de S\~ao Paulo (FAPESP). DC is
supported by DMR98-02373. Computations were performed at National
Computational Science Alliance (NCSA).

\end{document}